\documentclass[twoside,11pt]{article}

\usepackage{jsat}
\usepackage{amsmath, amsthm, amssymb}
\usepackage{pstricks}
\usepackage{pst-tree}

\input epsf

\jsatheading{??}{2013}{??-??}
\ShortHeadings{Reweighted BP for random $K$-SAT}
{F. Krzakala et al.}
\firstpageno{1}

\begin{document}

\title{Reweighted belief propagation and \\ quiet planting for random $K$-SAT}

\author{\name Florent Krzakala \email fk@espci.fr \\
	 \addr 
         Laboratoire de Physique Statistique, Ecole Normale Sup\'erieure, 45 rue d'Ulm\\
         Universi\'e P. et M. Curie, Paris, France \\
         ESPCI ParisTech, 10 rue Vauquelin, UMR 7083 Gulliver, Paris 75000, France.
       \AND
       \name Marc M\'ezard \email mezard@lptms.u-psud.fr \\
	 \addr Ecole Normale Sup\'erieure, 45 rue d'Ulm, Paris
         France. \\
        Universit\'e Paris-Sud \& CNRS, LPTMS, UMR8626, \\ B\^{a}t.~100, Universit\'e Paris-Sud 91405 Orsay, France. 
        \AND
       \name Lenka Zdeborov\'a \email lenka.zdeborova@cea.fr \\
	 \addr  Institut de Physique Th\'eorique, IPhT, \\ CEA Saclay,
         and URA 2306, CNRS,\\  91191 Gif-sur-Yvette, France.}
       
\maketitle

\begin{abstract}
 We study the random $K$-satisfiability problem using a partition function where each solution is reweighted according to the number of variables that satisfy every clause. We apply belief propagation and the related cavity method to the reweighted partition function. This allows us to obtain several new results on the properties of random $K$-satisfiability problem. In particular the reweighting allows to introduce a planted ensemble that generates instances that are, in some region of parameters, equivalent to random instances. We are hence able to generate at the same time a typical random SAT instance and one of its solutions. We study the relation between clustering and belief propagation fixed points and we give a direct evidence for the existence of purely entropic (rather than energetic) barriers between clusters in some region of parameters in the random $K$-satisfiability problem. We exhibit, in some large planted instances, solutions with a non-trivial whitening core; such solutions were known to exist but were so far never found on very large instances. Finally, we discuss algorithmic hardness of such planted instances and we determine a region of parameters in which planting leads to satisfiable benchmarks that, up to our knowledge, are the hardest known. 
\end{abstract}

%\keywords{SAT-solver, lookahead, equivalence reasoning, local learning}

%\published{September 2004}{October 2004}{November 2004}

\section{Motivation}

\def\(({\left(}
\def\)){\right)}                       
\def\[[{\left[}
\def\]]{\right]}
\def\e{{\rm e}}
\def\POP{{\rm POP}}

\newcommand{\<}{\langle}
\renewcommand{\>}{\rangle}
\newcommand{\be}{\begin{equation}}
\newcommand{\ee}{\end{equation}}
\newcommand{\bea}{\begin{eqnarray}}
\newcommand{\eea}{\end{eqnarray}}
\newcommand{\sign}{\text{sign}}
\renewcommand{\a}{\alpha}
\renewcommand{\b}{\beta}
\newcommand{\mc}{\mathcal}
\newcommand{\ox}{\overline{x}}
\newcommand{\s}{\sigma}
\newcommand{\uh}{\widehat{u}}
\newcommand{\Ch}{\widehat{C}}
\newcommand{\Fh}{\widehat{F}}
\newcommand{\Ph}{\widehat{P}}
\newcommand{\Qh}{\widehat{Q}}
\newcommand{\Q}{Q}
\newcommand{\deh}{\widehat{\delta}}
\newcommand{\de}{\delta}
\newcommand{\rh}{\widehat{\rho}}
\renewcommand{\r}{\rho}
\newcommand{\eh}{\widehat{\epsilon}}
\newcommand{\rr}{\widehat{r}}
\newcommand{\Th}{\widehat{T}}
\renewcommand{\th}{\widehat{t}}
\newcommand{\us}{\underline{\sigma}}
\newcommand{\da}{\partial a}
\newcommand{\di}{\partial i}
\newcommand{\ex}{\text{e}}

The satisfiability of Boolean formulas is a fundamental problem in theoretical computer science. It was the first problem shown to be NP-complete \cite{Cook71,levin1973universal,Papadimitriou94}, and it is of central relevance in various practical applications, including artificial intelligence, planning, hardware and electronic design, automation, verification and more. It can thus be thought of as the ``Ising model'' of computer science. Ensembles of randomly generated instances of the satisfiability problem have emerged in computer science as a way of evaluating algorithmic performance and addressing questions regarding the average case complexity.

An instance of the random ``$K$-SAT'' problem consists of $N$ Boolean variables and $M$ clauses.
Each clause contains a subset of $K$ distinct variables chosen uniformly at random, and each clause forbids one random assignment of these $K$ variables out of the $2^K$ possible ones. 
The instance is \textit{satisfiable} if there exists an assignment of variables that simultaneously satisfies all clauses. In this case we call such an assignment a {\it solution} to the instance. When the density of constraints, defined as $\alpha=M/N$, is increased, the formulas become less likely to be satisfiable. In the thermodynamic limit, i.e. when $N\to\infty$ at fixed $\alpha$, there is a sharp transition from a phase, $\alpha<\alpha_c(N)$, in which the formulas are almost surely satisfiable to a phase, $\alpha>\alpha_c(N)$, where they are almost surely unsatisfiable \cite{Friedgut99}. Although the convergence of the sequence $\alpha_c(N)$ as $N\to \infty$ is still an open problem, it is widely believed that this sequence is convergent, and therefore that there is a phase transition at $\alpha_c=\lim_{N\to\infty}\alpha_c(N)$. It is also a well known empirical result that the hardest instances are found near to this threshold \cite{CheesemanKanefsky91,MitchellSelman92,KirkpatrickSelman94}.

Random $K$-SAT has attracted the interest of mathematicians, computer scientists and statistical physicists. One very fruitful methodological direction to study random $K$-SAT is the belief propagation (BP) algorithm \cite{Pearl82,KschischangFrey01}, 
which is closely related to the cavity method that was developed in statistical physics for studies of mean field spin glasses \cite{MezardParisi87b,MezardParisi01}. The results and insights coming from the cavity method are remarkable. The satisfiability threshold and other phase transitions in the structure of solutions have been described in \cite{BiroliMonasson00,MezardZecchina02,MezardParisi02,KrzakalaMontanari06}.  In particular, it was shown that for $K\ge 3$ the space of solutions for highly constrained but still satisfiable instances splits into exponentially many (in $N$) clusters which are far away from each other.
Another important concept studied recently in random $K$-SAT \cite{Semerjian07,MontanariRicci08,ArdeliusZdeborova08} is the one of frozen variables: a given variable is frozen relatively to a given cluster of solutions if the variable is fixed to the same value in all the solutions of this cluster. Clearly any local search procedure which approaches a cluster must identify correctly the frozen variables.

The reader should note that, from a mathematical point of view, the majority of the results obtained with the cavity method are only conjectures since certain assumptions of the method were not proven yet. However, those assumptions are widely believed to be correct and hence these results, as well as the results of the present article, should be regarded as ``presumably-exact conjectures'' based on the so far heuristic, but sophisticated ``cavity method'', and the related ``replica method''. Notably a number of the predictions coming from this method (e.g. the existence of clustering of solutions) have been confirmed rigorously in random $K$-SAT \cite{MezardMora05,AchlioptasRicci06}. And an even larger number of these ``presumably-exact conjectures''  have been proven for other random constraint satisfaction problems such as coloring of random graphs, NAE-SAT or $K$-XOR-SAT since for those problems some methodological simplifications arise, see e.g. \cite{cojazdeborova2012,coja12,molloy12}.

In spite of the large corpus of results/conjectures obtained in recent years via the cavity method, there are still many open questions in random $K$-SAT, even on this heuristic level. Let us mention the main ones.
\begin{itemize}
\item
The relation between detailed properties of the energy landscape (where ``energy'' is defined as the number of unsatisfied clauses, and its domain are all the possible assignments) as predicted by the cavity method on one hand, and what is found by numerical simulations on the other hand is still not fully established. In the limit of large $K$, the situation is relatively well under control \cite{KrzakalaMontanari06,AchlioptasCoja-Oghlan08}:  phase transitions in the space of solutions can be identified with a threshold beyond which a large class of simple algorithms (or at least the methods used to analyze them) fail. However, for small $K$ the algorithmic limits are strongly algorithm-dependent, and many algorithmically-relevant questions about the energy landscape are still largely open. In particular, relatively simple stochastic-local-search algorithms have been found to be efficient in the regimes of the density of constraints where clustering of solutions occurs \cite{SeitzAlava05}. 
\item
Another annoying observation is that so far no one has been able to find solutions of large (consider e.g. $N>10^5$) random $K$-SAT formulas with frozen variables (i.e. variables which take the same value in all solutions) in spite of the fact that theory clearly predicts their existence \cite{MezardZecchina02,ZdeborovaKrzakala07} at large enough density of constraints. 
\item
Another aspect of this paradox is the fact that it was impossible so far to find numerically non-trivial solutions of the BP equations, while the theory shows that there are exponentially many of them (each associated with one cluster of solutions), and actually an algorithm like survey propagation is precisely based on the idea of counting the BP fixed points associated with frozen clusters \cite{MezardZecchina02}.
\end{itemize}

In this work we shall address some of these open issues by developing a `quiet planting' procedure for the random $K$-SAT problem. Planting is a way to generate a $K$-SAT instance with a known solution. Instead of generating the clauses randomly and then trying to find a solution, one first generates a special random configuration of variables, which we call the `planted configuration' and then one generates the clauses in such a way that each clause is satisfied by the planted configuration. By construction the planted configuration is a solution of the instance so generated, one says that it has been `planted' in the instance. Unfortunately, the naive planting that we have just described generates some instances that are very distinct from typical random $K$-SAT instances. A more careful procedure, called `quiet planting', allows to generate $K$-SAT instances that have the same distribution as fully random instances (when the density of constraints is below the threshold $\alpha_c$, to be defined later). Therefore quiet planting is a procedure that provides a random instance together with a solution. So far quiet planting had been developed for other constraint satisfaction problems \cite{AchlioptasCoja-Oghlan08,KrzakalaZdeborova09,KrzakalaZdeborova10,ZdeborovaKrzakala10}. In this paper we study quiet planting in  $K$-SAT. In order to do so, we introduce and study the {\it reweighted belief propagation} and the associated cavity method results. This study allows to answer several of the important open questions. From a mathematically rigorous point of view we provide new conjectures about the random $K$-SAT problem. More specifically, we address the following points: 
\begin{itemize} 
\item 
We find numerically some non-trivial BP  fixed points by initializing BP in the quietly-planted configuration. This shows that such non-trivial fixed points do exist, as expected from the cavity method analysis.  This answers a long-lasting question asked by random $K$-SAT practitioners who could not find these non-trivial BP fixed-points. They exist, but they are hard to find, and one needs a special procedure like quiet planting in order to see them.
\item 
Clusters are often imprecisely described in literature as connected components of the ``solution graph''. This is defined as a graph in the space of solutions where edges connect solutions that differ in no more than some number $d\ge 1$ of variables (the precise value of $d$ depends on the problem or on the author). Hence to walk from one such cluster to another one, it is necessary to violate a number of clauses $n(d)>0$ on the way, this is called an 'energetic' barrier between clusters.  It is known that clusters separated by energetic barriers exist (in the appropriate range of $\alpha$), when $K$ is large enough. 
Here we show the existence of different type of barriers, called ``entropic barriers''. These occur whenever a connected component of the solution graph is made of two subsets of solutions connected by a very narrow path of solutions: The path from one subset to another does not require violation of any clause, but this path is rare and a random walker would need a very long time, typically a time that diverges exponentially when $N\to \infty$.  Using the reweighted BP we explicitly show that entropic barriers do exist in random $K$-SAT. Entropic barriers had been known to exist in simple spin glass models (see for instance \cite{BarratMezard95,Ritort95,MoraZdeborova07}), but it is the first time that they are found in $K$-SAT.
\item An interesting problem which has been studied on several occasions in the literature \cite{BarthelHartmann02,MooreJia05} is: how to plant a solution in a $3$-SAT problem such that the resulting formula is nevertheless very hard. In this paper we clarify in which region the planted instances are hard, namely in the shaded region in Fig.~\ref{fig1}. Qualitatively we reach the same conclusions as \cite{BarthelHartmann02}. We give the exact boundary of the hard region (whereas the calculations in \cite{BarthelHartmann02} were only approximate) and we provide more detailed heuristic arguments for this result. In the end we have a way to create the hardest known (at least to the authors) satisfiable $3$-SAT formulas. Note that the hard ensemble discussed here is closely related to the one of \cite{HaanpaaJarvisalo06}. 
\end{itemize}

\section{Reweighting in random $K$-SAT}
The satisfiability problem is defined over $N$ Boolean variables $s_i\in\{0,1\}$, and $M$ clauses. Each clause contains $K$ variables. If a variable $i$ is negated in the clause $a$ we set $J_{ia}=1$, otherwise we set $J_{ia}=0$.  A clause $a$ is satisfied if and only if 
\be
n_a =  \sum_{i\in \partial a} (1-\delta_{s_i,J_{ia}})>0\, ,
\ee 
where $\partial a$ is the set of variables belonging to the clause $a$. Here we introduced $n_a$ as the number of variables that are satisfying the clause $a$.

Let us first define the standard probability measure over solutions. We assume that there exists at least one solution, and we define the partition function as the number of solutions:
\be
    Z  = \sum_{\{s_i\}}   \prod_{a} C_a(\{s_i\}_{i\in \partial a})\, ,
\ee
where $C_a=1$ if clause $a$ is satisfied and $C_a=0$ otherwise. The sum is over all the $2^N$ assignments of variables. One introduces a `Boltzmann'-type measure as the uniform measure over all solutions:
 \be
     \mu(\{s_i\})  = \frac{1}{Z} \prod_{a} C_a(\{s_i\}_{i\in \partial a})\, .
\ee
The cavity method (or BP) can be used to study heuristically the properties of this `standard'  measure.

Let us now define a  reweighted measure. We introduce $K$ parameters $\lambda_1,\dots,\lambda_K$ which are non-negative real numbers, and we define the `$\lambda$-measure' as:
\be
 \mu_\lambda(\{s_i\})  = \frac{1}{Z(\lambda)} \prod_{a} \left[ C_a(\{s_i\}_{i\in \partial a}) \, \lambda_{n_a(\{s_i\}_{i\in \partial a})} \right] \, , \label{rev_Boltz}
\ee
Where the reweighted partition function is defined as
\be
   Z(\lambda)  = \sum_{\{s_i\}}   \prod_{a} \left[ C_a(\{s_i\}_{i\in \partial a}) \, \lambda_{n_a(\{s_i\}_{i\in \partial a})} \right] \, . \label{rev_Z}
\ee

We denote the vector of $\lambda=(\lambda_1,\lambda_2,\dots, \lambda_K)$. 
In order to have a unique measure associated with each vectors $ \lambda$, we choose to fix the normalization
\be
     1 =  \sum_{r=1}^K {K\choose r} \lambda_r \, .  \label{norm}
\ee 
(note that any normalization procedure that breaks the symmetry of (\ref{rev_Boltz}) with respect to a global multiplication of all components of $\lambda$ would be fine, we chose eq.~(\ref{norm}) because it leads to convenient simplifications in some expressions that follow). 
The standard measure and partition function are recovered when all components of $ \lambda$ are equal: $\forall r \ \lambda_r=1/(2^K-1)$.

The main subject of this paper is to generalize the cavity-method-based conjectures on the phase diagram of random $K$-SAT to the generalized reweighted $K$-SAT problem  with  vector-parameter $ \lambda$, and to discuss new interesting results which can be derived using the generalized measure (\ref{rev_Boltz}).
There are some properties that do not depend on the reweighting parameters $ \lambda$ (as long as $\lambda_r>0$ for all $r=1,\dots,K$), for instance the satisfiability threshold or the freezing transition, which is the density of constraints at which strictly all clusters start to contain frozen variables. But other physically important phase transition defined in \cite{KrzakalaMontanari06,MontanariRicci08} such as the clustering or the condensation transitions do depend on reweighting parameters.

Note that special cases of the reweighted partition function were studied previously. In particular, \cite{AchlioptasPeres04} used the set of reweighting parameters
\be
\lambda_r = \frac{\gamma^r}{ \sum_{r=1}^K {K\choose r} \gamma^r } \label{lambda_power} 
\ee 
as a tool to be able to use the second moment method to obtain a lower bound on the satisfiability threshold. Indeed, the success of this reweighting in the second moment calculations inspired our study. 
%In fact, in the region where a second moment lower bound can be proven, the quenched and annealed entropy densities are equal and this is a condition for the quiet planting to work. 
The same reweighting was later used in several other rigorous works in conjunction with the second moment method. Let us also mention the work of \cite{MooreJia05} where planting according to the reweighting (\ref{lambda_power}) was studied numerically, together with the first and second moments calculations. However, it seems that the authors of \cite{MooreJia05} did not notice that for a special value of $\gamma$ the generated instances are equivalent to random instances, which is one of the crucial points in the present paper. Reweighted planting was also studied in \cite{BarthelHartmann02} as a way to create hard instances. We shall comment on all these results later in the light of our findings.

\section{Belief propagation for the reweighted partition function}
The reweighted partition function (\ref{rev_Z}) can be computed via the belief propagation (BP) algorithm \cite{Pearl82,KschischangFrey01}.  The BP result is exact on trees, and it corresponds to the Bethe approximation on sparse random graphs. Considering the probability measure (\ref{rev_Boltz}), we define $\psi_{s_i}^{a\to i}$ as the probability that the constraint $a$ is satisfied, conditioned on the fact that the value of  variable $i$ is $s_i$. Similarly,  $\chi_{s_j}^{j\to a}$ is the probability that the variable $j$ takes value $s_j$ conditioned on the fact that the constraint $a$ has been removed from the graph. According to belief propagation these messages then satisfy the set of equations
\bea
    \chi_{s_i}^{i\to a} &=& \frac{\prod_{b\in \partial i \setminus a}  \psi_{s_i}^{b\to i} }{\prod_{b\in \partial i \setminus a}  \psi_{1}^{b\to i}+\prod_{b\in \partial i \setminus a}  \psi_{0}^{b\to i}}\, , \label{BP_1} \\
     \psi_{J_{ia}}^{a\to i} &=& \frac{1}{Z^{a\to i}} \sum_{\{s_j\}_{j\in \partial a \setminus i}}
\theta (K-1-\sum_{j\in \partial a \setminus i} \delta_{J_{ja},s_j}) \, \lambda_{K-1 - \sum_{j\in \partial a \setminus i} \delta_{s_j,J_{ja}} } \prod_{j\in \partial a \setminus i} \chi_{s_j}^{j\to a}\, , \label{BP_2a} \\
     \psi_{1-J_{ia}}^{a\to i} &=& \frac{1}{Z^{a\to i}} \sum_{\{s_j\}_{j\in \partial a \setminus i}}  \lambda_{K-\sum_{j\in \partial a \setminus i} \delta_{s_j,J_{ja}} } \prod_{j\in \partial a \setminus i} \chi_{s_j}^{j\to a}\, ,\label{BP_2b} 
\eea 
where the normalization $Z^{a\to i}$ ensures that $\psi_{J_{ia}}^{a\to i} +  \psi_{1- J_{ia}}^{a\to i} =1$, and $\theta(x)$ is the Heaviside step function.
% A technical remark on the implementation of the BP equations, it may seem that the sum in eqs. (\ref{BP_2a}-\ref{BP_2b}) takes $2^{K-1}$ steps, which may be very costly. Note, however, that the terms in the sum depend only of the $\sum_{j}\delta_{J_{aj},s_j}$ hence using a convolution the sum can be computed iteratively in only $K^2$ steps.
 Once these messages (i.e. the solution of these equations) have been found, the marginal probability that variable $i$ takes value $s_i$, denoted by $\chi_{s_i}^{i}$, can be computed as:
\be
     \chi_{s_i}^{i} = \frac{\prod_{b\in \partial i }  \psi_{s_i}^{b\to i} }{\prod_{b\in \partial i }  \psi_{1}^{b\to i}+\prod_{b\in \partial i }  \psi_{0}^{b\to i}}\, .
\ee
The log-partition function can then be computed using the Bethe formula. This gives  the ``free-entropy'' density:
\be
     s(\lambda)=\frac{1}{N}\log{Z(\lambda)} =\frac{1}{N}\left( \sum_{a} \log{Z^a} + \sum_i \log{Z^i} - \sum_{ia} \log{Z^{ia}}\right)\, , \label{Bethe_ent}
\ee
where 
\bea
      Z^i &=&  \prod_{a\in \partial i }  \psi_{1}^{a\to i}+\prod_{a\in \partial i }  \psi_{0}^{a\to i}\, ,  \\
       Z^a &=&  \sum_{\{s_i\}_{i\in \partial a }} \theta(K-\sum_{i\in \partial a} \delta_{J_{ia},s_i} ) \, \lambda_{K - \sum_{i\in \partial a} \delta_{s_i,J_{ia}} } \prod_{i\in \partial a } \chi_{s_i}^{i\to a} \, ,\\
    Z^{ia} &=& \chi_{0}^{i\to a} \psi_{0}^{a\to i} + \chi_{1}^{i\to a} \psi_{1}^{a\to i} \, .
\eea

\section{Quiet planting in random $K$-SAT}

A reweighting parameter set $ \lambda^*$ is defined as ``magic'' when the set of eqs. (\ref{BP_1}-\ref{BP_2b}) has a solution (fixed point) for which the messages do not depend on the edge $(ai)$, i.e. $\chi^{i\to a}_s = \chi_s$, and $\psi^{a\to i}_s=\psi_s$ for all $(ai)$. Such a solution (fixed point) is referred to in the statistical physics literature as the ``factorized" fixed point of belief propagation. We shall see that problems defined with magic reweigthing have special properties.

\subsection{Magic reweigthing and quenched-annealed equivalence}
Since in random $K$-SAT the negations are chosen at random there is a global symmetry between 0's and 1's. A factorized fixed point of BP hence also has to be symmetric, i.e. $\psi_{s}^{a\to i}=\chi_{s}^{i\to a}= 1/2$ for all $(ai)$ and $s$. With the use of normalization (\ref{norm}) it is easy to see that Eqs.~(\ref{BP_2a})-(\ref{BP_2b}) permit such a fixed point if and only if $\lambda$ satisfies
\be
      \frac{1}{2} = \sum_{r=1}^K {K-1 \choose r-1} \lambda^*_r  \, . \label{lambda_q}
\ee
Note that, if we restrict to the power-law choice of the reweighting parameters (\ref{lambda_power}) made in \cite{AchlioptasPeres04,MooreJia05}, we recover from (\ref{lambda_q}) the condition for the special value $\gamma=\gamma^*$ used in these works, defined by:
\be
  1 = (1+\gamma^*)^{K-1}(1-\gamma^*)\, .
\label{gamma_star}
\ee 
In what follows we will mostly speak about the general $\lambda^*$ magic reweighting, but sometimes we focus for simplicity on the power-law choice of $\gamma^*$. We urge the reader not to confuse the two, as $\gamma^*$ is only one special point in the space of magic reweigthings. 

The free-entropy density (\ref{Bethe_ent}) in the case of a factorized BP fixed point is equal to 
\be
      s_{\rm fBP}(\lambda^*) = (1-\alpha K)  \log{2}\, , \label{s_para} 
\ee
this is obtained simply by evaluating eq. (\ref{Bethe_ent})  for $\psi_{s}^{a\to i}=\chi_{s}^{i\to a}= 1/2$.

We now define the quenched and annealed free-entropy density as
\bea
s_{\rm annealed}(\lambda)&\equiv& \lim_{N\to \infty} \frac{1}{N}\log\mathbb{E}[{Z(\lambda)}]\, , \\ 
s_{\rm quenched}(\lambda)&\equiv&\lim_{N\to \infty}\frac{1}{N}\mathbb{E}[\log{Z(\lambda)}]\, .
\eea
where $Z(\lambda)$ is the partition function defined in (\ref{rev_Z}) and the expectation is with respect to the measure (\ref{rev_Boltz}). These names are motivated from properties of glassy material after a fast (quenched) or slow (annealed) decrease of temperature.  It follows from the cavity method that under certain conditions the Bethe free-entropy density of the factorized fixed point is equal to the quenched free-entropy density $s_{\rm fBP}(\lambda^*) = s_{\rm quenched}(\lambda^*)$ in the thermodynamic limit $N\to\infty$. The condition for this equality to hold is that the density of constraints, $\alpha$ should be smaller than a threshold value where a thermodynamic phase transition called the ``condensation transition'' occurs. This transition is  defined by the non-analyticity of the quenched free-entropy.

Let us now compute the annealed free-entropy density of the reweighted ensemble with arbitrary $\lambda$. We get
\be
       \mathbb{E}[{Z(\lambda)}] = 2^N \left[ \sum_{r=1}^K \lambda_r {K\choose r} \frac{1}{2^K}\right]^{\alpha N}  = 2^N 2^{-\alpha K N}\, ,
\ee
where in the second equality we used the normalization condition (\ref{norm}). We see that  the annealed free-entropy density equals the Bethe free-entropy density of the factorized BP solution (\ref{s_para}) found for $\lambda=\lambda^*$, $s_{\rm annealed}=s_{fBP}(\lambda^*)$. This is actually a general property, see \cite{Mora07}. 
Consequently, at low enough density of constraints,  the quenched free-entropy density at $\lambda^*$ is equal to the annealed free-entropy density. This last property is the crucial point that makes quiet planting in the sense of \cite{AchlioptasCoja-Oghlan08,KrzakalaZdeborova10,ZdeborovaKrzakala10} possible. 

\subsection{Quiet planting in the reweigthed ensemble}

The definition of the $\lambda$-reweighted planted ensemble is the following:
\begin{enumerate}
\item
 Choose the geometry of the random $K$-SAT formula as before ($M$ clauses where each contains a uniformly-chosen random $K$-tuple of variables).
\item
Define a special configuration, called the  {\it planted} configuration, by assigning randomly  1/2 of the variables to the value $0$ and the other 1/2 to $1$. 
\item
Define the probability $p(r)$ by: 
\be
    p(r) = {K\choose r} \lambda_r \quad \quad {\rm for} \quad r>0\, ,\quad p(0)=0\, . \label{p_t}
\ee
For each clause $a$,  choose a random integer $r_a>0$ from this distribution, and then choose at random one out of the ${K\choose r_a }$ configurations of negations for which the planted configuration satisfies the clause by $r_a$ different variables.  
\end{enumerate}
In this planted ensemble, the probability  that a randomly chosen clause is satisfied by $r$ variables from the planted configuration is thus equal to $p(r)$ defined in (\ref{p_t}).

First of all notice that since every planted configuration is consistent with the same number of instances the planted configuration has all the properties of a configuration sampled uniformly at random from the corresponding probability measure $\mu(\lambda)$, eq. (\ref{rev_Boltz}). This is a very generic property of planting.
Furthermore, it is easy to see that this planting procedure generates instances randomly but in general not uniformly over all instances. Every planted instance appears with probability proportional to its partition function $Z(\lambda)$. Hence the planted ensemble of instances is in general very different from the random ensemble. Only in the case when the annealed free-entropy density equals the quenched one, the fluctuations of $Z(\lambda)$ are small enough (in the large $N$ limit) so that the planting procedure generates instances statistically equivalent to the random ones. This property called ``quiet planting'' means that every property that holds with high probability in the random ensemble holds also with high probability in the planted ensemble and vice versa,  as discussed in \cite{AchlioptasCoja-Oghlan08,KrzakalaZdeborova10,ZdeborovaKrzakala09,MosselNeeman12}. 

To summarize, when one uses magic-reweighting, with constraint densities smaller than the corresponding condensation phase transition \cite{KrzakalaMontanari06}, the annealed free-entropy density equals the quenched one and the planting is quiet. This means that the planting procedure has the following two properties: First it creates formulas of the satisfiability problem that are statistically equivalent to random formulas in the thermodynamic limit $N\to \infty$. Second, the planted configuration has all the properties of a configuration sampled randomly from the measure $\mu(\lambda^*)$. In physics language, the planted configuration is an ``equilibrium configuration'' of the measure $\mu(\lambda^*)$ (while it is not an equilibrium configuration of the uniform measure over all solutions). In the next sections we explore consequences of these simple but very useful properties. 

\subsection{Cavity equations in the ``magically" reweighted ensemble}
Iterating the BP equations on a given instance generated with quiet planting until a fixed point is reached is easy and provides an experimental access to the study of phase transitions in this problem. However, the very notion of phase transition is an asymptotic one (it exists only in the thermodynamic limit $N\to\infty$), and it is thus very useful to get a control of this thermodynamic limit. This is done in the cavity method by studying an ensemble of planted problems. The corresponding distributional equations can be solved efficiently with population dynamics technique \cite{MezardParisi01}, these equations are referred to as the replica-symmetric cavity equations. 

In the present case the object studied by the cavity method is the probability  $P_{s_i}^{t}(\psi)$  that at the $t$-th iteration of BP equations a message arriving on a site $i$ takes the value $\psi=(\psi_0,\psi_1)$,
conditionally on the fact that the planted configuration on $i$ is equal to $s_i$.
It satisfies:
\bea
    P_{s_i}^{t} (\psi) & = & \sum_{J_j} \prod_{j=1}^{K-1} \frac{\delta_{J_j,0}+\delta_{J_j,1}}{2} \sum_{\{s_j\}}   \tilde P(\{s_j\},s_i)  \nonumber \\  & & \int \prod_{j=1}^{K-1} \sum_{l_j=1}^\infty q(l_j) \prod_{k_j=1}^{l_j} P_{s_j}^{t-1} (\psi^{k_j})\,  {\rm d}\psi^{k_j} \delta(\psi - {\cal F}(\{\psi^{k_j}\})) \label{pop_dyn}
\eea
where the function ${\cal F}(\{\psi^{k_j}\})$ is obtained by combining (\ref{BP_2a}-\ref{BP_2b}) with (\ref{BP_1})
\bea
     {\cal F}_{J_{i}}(\{\psi^{k_j}\})  &=& \frac{1}{\tilde Z} \sum_{\{s_j\}_{j=1}^{K-1}}
\theta (K-1-\sum_{j=1}^{K-1} \delta_{J_{j},s_j}) \, \lambda_{K-1 - \sum_{j=1}^{K-1} \delta_{s_j,J_{j}} } \prod_{j=1}^{K-1} \prod_{k_j=1}^{l_j}  \psi_{s_j}^{k_j} \, , \label{BP_1pop} \\
     {\cal F}_{1-J_{i}}(\{\psi^{k_j}\})   &=& \frac{1}{\tilde Z} \sum_{\{s_j\}_{j=1}^{K-1}}  \lambda_{K-\sum_{j=1}^{K-1} \delta_{s_j,J_{j}} } \prod_{j=1}^{K-1} \prod_{k_j=1}^{l_j}  \psi_{s_j}^{k_j} \, ,\label{BP_2pop} 
\eea
and $q(l_j)$ is the Poisson distribution with mean $\alpha K$. The probability distribution $\tilde P(\{s_j\},s_i)$ is obtained by drawing a random variable $J_i$ uniformly from $\{0,1\}$ and then a number $r$ of variables $s_j$'s such that $s_j\neq J_j$, where $r$ is drawn from the distribution $\tilde p(r)$:
\bea
     \tilde p(r) = \frac{rp(r)}{\sum_{r'=0}^K r' p(r')}  \quad {\rm if}\quad  J_i \neq s_i \, , \\
     \tilde p(r) =  \frac{(K-r)p(r)}{\sum_{r'=0}^K (K-r') p(r')}\quad   {\rm if} \quad  J_i = s_i \, . \label{p_tt}
\eea
Where $p(r)$ is given by (\ref{p_t}). The planted initialization then corresponds to $P^{\rm init}_{s_i}(\psi)=\delta(\psi -  u)$ where $u_{s_i}=1$ and $u_{1- s_i} =0$. The average Bethe free-entropy is expressed along the same lines. 

\section{Phase transitions in the reweighted ensemble, or how to use quiet planting}
Following the argumentation in \cite{KrzakalaZdeborova09}, the quietly-planted satisfiability instances have in general three sharp phase transitions as the density of constraints increases. The simplest way to locate these phase transitions is to directly perform the planting, and then iterate till convergence the reweighted BP equations (\ref{BP_1}-\ref{BP_2b}) on the resulting instance, with two possible initializations:
\begin{itemize}
\item
``Random initialization''. One initializes messages $\chi^{i\to a}$ and $\psi^{a\to i}$ as random vectors drawn uniformly from the set of normalized vectors.
In the cavity equations, $P_{s_i}^{0}(\psi)$ is the uniform density over the set $\psi_0+\psi_1=1$.
\item
``Initialization in the planted configuration''. One initializes the messages in the planted configuration, i.e. $\chi^{i\to a}_{1} = 1$, $\chi^{i\to a}_0=0$ if variable $i$ was planted $1$ or $\chi^{i\to a}_{1} = 0$, $\chi^{i\to a}_0=1$ if variable $i$ was planted $0$. The distribution  $P_{s_i}^{0}(\psi)$  in the cavity equations is the one induced by this choice, using Eqs.~(\ref{BP_2a}-\ref{BP_2b}).
\end{itemize}

Let us summarize the general scenario found in quiet planting. The following results can be found empirically by performing the numerical experiment of iterating BP equations on large instances (which is easily done), but the real control of the phase transitions is done through the 
asymptotic analysis of the distributional equations of the cavity method written in the previous section.

In order to describe the various phases of the system, we use a vocabulary borrowed from statistical physics.
We say that a system is ``ferromagnetic''  when the expectation of a variable has a positive correlation with the planted configuration (in the same way that a system of spins is ferromagnetic if the variables -the spins - have a positive correlation with a given spatial orientation, the orientation of the magnetic moment of the material).
If we call  $\tau$ the planted configuration, the probability measure $\mu(\{s_i\})$ is ferromagnetic if in expectation the configurations sampled from the measure have strictly positive overlap with the planted configuration, where the overlap $q$ is defined as:  $q= \lim_{N\to\infty}(1/N)\sum_{i=1}^N [\mu(s_i=\tau_i) -\mu(s_i \neq \tau_i)]>0$. If this overlap is zero we call the measure ``paramagnetic''. 

As a function of the constraint density $\alpha$ we find four different phases in the quietly-planted satisfiability, identified by the behavior of the fixed points obtained from iterating BP with our two different initial conditions.
\begin{itemize}
  \item{For $\alpha<\alpha_d$ the system is in a purely paramagnetic phase. BP converges towards the factorized fixed point when started from the two initializations. The overlap between the BP fixed point and the planted configuration is zero. It turns out that $\alpha_d$ is the clustering threshold. In this phase the set of solutions forms one cluster, two random configurations from this cluster are completely uncorrelated, the planted configuration is simply one of them, and there is no way to tell which one.  }
  \item{For $\alpha_d<\alpha<\alpha_c$ there is a paramagnetic phase with a subdominant (metastable) ferromagnet. BP started from the random initialization converges to the factorized fixed point, whereas BP started from the planted initialization converges to a fixed point with positive overlap with the planted configuration. The Bethe free-entropy of this second fixed point is smaller than the annealed free-entropy and hence the paramagnet is still the dominant solution. The Bethe free-entropy of the factorized fixed point is still equal to the quenched free-entropy. 
It turns out that $\alpha_c$ is the condensation threshold. In this phase there are exponentially many clusters of roughly the same size, the planted configuration belongs to one of them and there is no way to tell which one it belongs to. } 
  \item{For $\alpha_c<\alpha<\alpha_l$} there is a ferromagnetic phase with a subdominant (metastable) paramagnet. Like in the previous phase, BP started from the random initialization still converges to the factorized fixed point, whereas BP started from the planted initialization converges to a fixed point with positive overlap with the planted configuration. 
But this time the Bethe free-entropy of the BP fixed point obtained from the planted initialization is larger than the Bethe free-entropy of the factorized fixed point. Hence the planted cluster dominates the measure. It is now the Bethe free-entropy of the BP fixed point obtained from the planted initialization that is equal to the quenched free-entropy.  In this phase the planting is not quiet anymore, in the sense that the properties of planted and random instances are different. For instance random instances become typically unsatisfiable for $\alpha>\alpha_s>\alpha_c$ ($\alpha_s$ being the satisfiability threshold), whereas the planted ones are always satisfiable. 
  \item{For $\alpha_l<\alpha$} there is a ferromagnetic phase. In this phase the two initializations give the same result: in both cases  BP converges to a fixed point which is correlated with the planted configuration. Hence it is easy to find a solution correlated with the planted configuration. 
\end{itemize}

We emphasize here that the planting is ``quiet'' only for $\alpha \le \alpha_c$, whereas the properties of the random and planted ensemble are different for $\alpha>\alpha_c$.

%ICICI
In the above discussion the phases have been identified by analyzing the behavior of BP fixed-points. It turns out that the phase transitions found in this way also correspond to phase transitions in the geometrical properties of the space of solutions of satisfiability found in previous works. We shall state this correspondance here without proof.

In the random $K$-SAT ensemble, reweighted with the parameter $\lambda^*$, $\alpha_d$ is the threshold that corresponds to  the clustering (called dynamical 1RSB in statistical physics) transition of the measure $\mu(\lambda^*)$. For $\alpha<\alpha_d$ the set of solutions form one cluster. The planted configuration is one configuration of this cluster, it has no special property.
The threshold  $\alpha_c$  is the condensation (called static 1RSB in statistical physics) transition of the measure $\mu(\lambda^*)$. 
The phase transition at $\alpha_l$ is a point beyond which the paramagnetic phase ceases to exist. In the language of magnetic systems, $\alpha_l$ is called the spinodal threshold.

Another important property that follows from the analysis of quiet planting \cite{KrzakalaZdeborova09} is that the part of the phase space that is not correlated with the planted configuration remains unchanged in the whole region of $\alpha$, i.e. also for $\alpha>\alpha_c$. Notably for $\alpha>\alpha_s$ (where $\alpha_s$ is the satisfiability threshold in the random SAT ensemble)
only solutions correlated to the planted configuration exist in the planted ensemble.  In the random $K$-SAT ensemble without planting, $\alpha_l$ does not have a physical interpretation, but at this point the BP iterations stop to converge. 

The transition point $\alpha_l$ is investigated by computing the local stability of the factorized BP fixed point with respect to small random perturbations. This is done by linearizing equations (\ref{BP_1}-\ref{BP_2b}) via expansion around the factorized fixed point and analyzing the largest eigenvalue $x$ of the associated linear operator. The following analytical formula then locates the phase transition between a phase in which an infinitesimal random perturbation is damped to zero from a phase where this perturbation gets amplified in the direction of the planted configuration: 
\be
   \alpha_l = \frac{1}{K(K-1) x^2} \quad {\rm where} \quad  x = \frac{ \sum_{r=1}^{K-2} {K-2 \choose r} \lambda_r  }{\sum_{r=1}^{K-1} {K-1 \choose r} \lambda_r } + \frac{\sum_{r=2}^{K} {K-2 \choose r-2} \lambda_r }{\sum_{r=1}^{K} {K-1 \choose r-1} \lambda_r } -1
% x = \frac{\lambda_q}{1+\lambda_q} + \frac{(\lambda_q+1)^{K-2}-1}{(\lambda_q+1)^{K-1}-1} -1 \, .
\ee
In order to locate the other phase transitions, $\alpha_d$ and $\alpha_c$,  we have solved the cavity equations (\ref{pop_dyn})-(\ref{p_tt}) using the population dynamics \cite{MezardParisi01}.  
%used the quiet planting described above and we have run the BP equations on large instances. Thanks to the  self-averaging property, this turns out to be an efficient way to get  neat results, using moderate computational resources. 

\begin{figure}[!ht]
\begin{center} 
\includegraphics[scale=0.6]{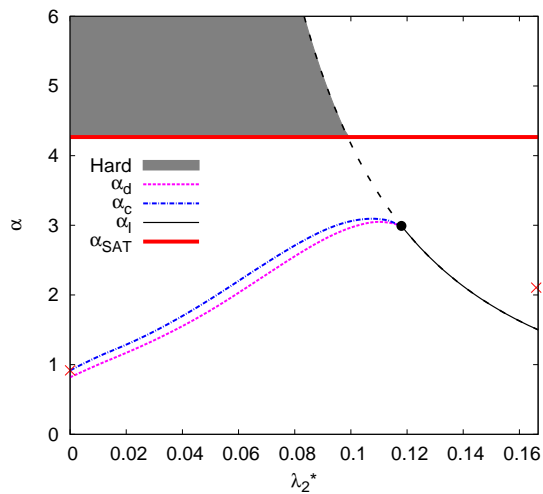} \caption{The 
phase diagram of quietly-planted reweighted 3-SAT in the plane of the reweighting parameter $\lambda^*_2$ and the density of constraints $\alpha$. We remind that the other two reweighting parameters  $\lambda^*_1$ and $\lambda^*_3$ are obtained 
using conditions (\ref{norm}) and (\ref{lambda_q}). Note that the boundary 
value $\lambda^*_2=0$ in some sense reduces the problem into planted 
3-XOR-SAT, and $\lambda_2^*=1/6$ to the planted NAE-SAT. Both these cases
 were studied in \cite{MezardRicci03,CastellaniNapolano03} and phase transitions
 in these cases were known (the SAT/UNSAT transition for these problems is shown 
as a red cross). The SAT/UNSAT transition does  not depend on $\lambda_2^*$, 
except for these two boundary cases. The black dot corresponds to the phase transition for the power-law reweighted ensemble
 with parameter $\gamma^*$, which is indistinguishable within our accuracy from 
the point where the transition goes from a first order (dash-dotted blue line) to a second order (full black line) one. The curve $\alpha_l$ is dashed in the region where it corresponds to a spinodal of the first order phase transition and not the second order phase transition. The shaded
 region marks the part of the phase diagram where planting with the corresponding $\lambda^*$ creates extremely hard (as will be quantified in section \ref{sec:hard}) 3-SAT satisfiable instances. }  \label{fig0} 
\end{center}
\end{figure}

In Fig.~\ref{fig0} we plot the full phase diagram for quietly planted 3-SAT. In this case the vector $\lambda^*$ has only one free parameter, we chose it to be $\lambda_2$, the other two components are then given from conditions (\ref{norm}) and (\ref{lambda_q}). The phase transition is continuous, i.e. $\alpha_d=\alpha_c=\alpha_l$, for $\lambda_2^*>0.118(3)$. It is discontinuous, i.e. $\alpha_d<\alpha_c<\alpha_l$,  for $\lambda_2^*$ smaller than this threshold. Within our numerical precision this tri-critical point in random 3-SAT agrees with the value of $\lambda_2^*=0.1180$ corresponding to the $\gamma^*$ reweighting (\ref{lambda_power}). We did not find a theoretical reason for this.

\begin{figure}[!ht]
\centering  
 \includegraphics[scale=0.6]{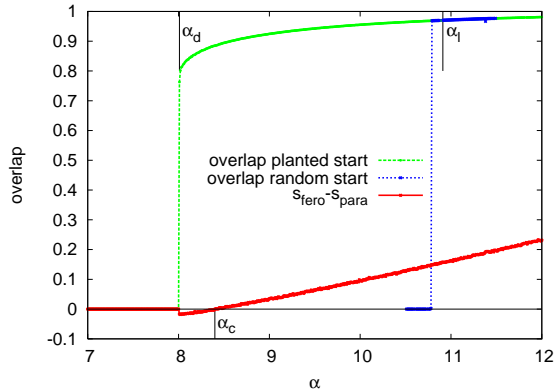} 
\caption{Phase diagram of the $\gamma^*$ planted $4$-SAT, obtained by the population dynamics solution of the cavity equations, population size was $5\cdot 10^4$. The overlap between the planted configurations and the BP fixed point when BP is initialized a) randomly - blue curve - showing a phase transition at $\alpha_d$,  b) in the planted configuration - green curve - showing a phase transition at $\alpha_l$. A careful reader will remark that the blue curve jumps up in the overlap a bit before the constraint density $\alpha_l$, this is dues to a strong finite size correction associated to this phase transition. The red curve is the difference between the paramagnetic free-entropy (\ref{s_para}) and the free-entropy corresponding to the BP fixed point obtained from the planted initialization. It becomes positive at the first order phase transition $\alpha=\alpha_c$: for $\alpha<\alpha_c$ the paramagnetic phase is the dominant one, for $\alpha>\alpha_c$ the ferromagnetic phase is the dominant one.}   
\label{fig1}  
\end{figure}

For the general case of $K>3$ we keep for simplicity to the power-law reweighting of  (\ref{lambda_power}) with the condition (\ref{gamma_star}), as it was used in \cite{AchlioptasPeres04,MooreJia05}.
Fig.~\ref{fig1} shows, in the case $K=4$, the properties of the BP-fixed points, starting from the two possible initial conditions.
The overlap of the resulting fixed points with the planted configurations allows to find the threshold $\alpha_d$. The difference of the Bethe entropies of the paramagnetic and ferromagnetic fixed point allows to locate the first-order phase transition $\alpha_c$.

Table \ref{table1} summarizes the values of these phase transitions for different values of $K$.  For comparison we also give the dynamical and condensation phase transitions in the canonical non-reweighted $K$-SAT that corresponds to $\hat \lambda_r=1/(2^K-1)$ for all $r=1,\dots,K$. 
\begin{table}[!ht]
\begin{center}
\begin{tabular}{|l||l||l|l|l|l||l|l||l|l|l|} \hline
K      & $\alpha_s$ & $\gamma^*$   &\, $\alpha_d^{\gamma^*}$\, &\, $\alpha_c^{\gamma^*}$&\, $\alpha_l^{\gamma^*}$&\, $\alpha_d^{\hat \lambda}$&\, $\alpha_c^{\hat \lambda}$ &\,  $\alpha_{BP(\hat \lambda)}^{\gamma^*}$ &\, $\alpha_{\rm 2nd}^{\gamma^*}$ &\, $\alpha_f^{\gamma^*}$\\ \hline 
3 & 4.27 &0.61803    & 2.991  & 2.991    & 2.991    & 3.86     & 3.86       & 3.93  & 2.574  & 4.29 \\ \hline
4 & 9.93 &0.83929    & 8.01   & 8.40     & 10.91    & 9.38     & 9.547      & 8.75  & 7.313  &  9.63 \\ \hline
5 & 21.12 &0.92756    & 17.2   & 19.30    & 35.40    & 19.16    & 20.80      & 17.7  & 17.617 & 19.4 \\ \hline
6 & 43.4 &0.96595    & 34.5   & 41.3     & 111.1    & 36.53    & 43.08      & 34.7  & 39.026 &  37.7 \\ \hline
%7 & 0.98358    & ?      & ?        & 347.6    & ?        & ?          &        \\ \hline
\end{tabular}  
\caption{\label{table1} For various values of $K$ we give the satisfiability threshold $\alpha_s$ \cite{MertensMezard06}, the ``quiet" value of the parameter $\gamma^*$, the corresponding dynamical, $\alpha_d^{\gamma^*}$, condensation, $\alpha_c^{\gamma^*}$, and spinodal, $\alpha_l^{\gamma^*}$, thresholds for the reweighted probability measure $\mu(\gamma^*)$ in random $K$-SAT. Recall that quiet planting generates typical random graphs for density of constraints smaller than $\alpha_c^{\gamma^*}$.  For comparison we remind the dynamical and condensation thresholds for the $K$-SAT with no reweighting, i.e. for the measure $\mu(\hat \lambda)$, from \cite{KrzakalaMontanari06}. 
The column $\alpha_{BP(\hat \lambda)}^{\gamma^*}$ gives the density of constraints beyond which even canonical BP (at $\hat\lambda$) converges to a non-trivial fixed point if initialized in the planted configuration. 
The $\alpha_{\rm 2nd}^{\gamma^*}$ column is taken from \cite{MooreMertens11,AchlioptasPeres04}, it is the rigorous lower bound on the satisfiability threshold that is obtained by computation of the reweighted second moment.  The work of \cite{MooreMertens11,AchlioptasPeres04} also implies that the annealed free-entropy density equals the quenched one in the range $\alpha < \alpha_{\rm 2nd}^{\gamma^*}$.  Our arguments lead us to conjecture that actually the annealed free-entropy density equals the quenched one in the whole range $\alpha\le \alpha_c^{\gamma^*}$. The last column $\alpha_f^{\gamma^*}$ is the constraint density beyond which the planted solution lies in a frozen cluster.}
\end{center}
\end{table}

\section{Results unveiled by quiet planting}

\subsection{Non-trivial whitening and BP fixed points}
In the standard $K$-SAT problem, without reweighting or planting, the iteration of BP equations starting from a generic random initial condition typically have two possible behaviors, depending on the value of $\alpha$. Either they always converge to the same fixed-point, that we call the ``trivial'' fixed point, or they do not converge at all (as in 3-SAT for $\alpha>3.86$).  On the other hand, the statistical physics description of the clustered phase starts from the assumption that there should exist one different and non-trivial BP fixed point  for each cluster of solutions, and then counts the number of non-trivial BP fixed points via an augmented-BP-like approach. This counting finds an exponentially large number of BP fixed points, indicating an exponentially large number of clusters \cite{MezardZecchina02,MezardMontanari09}. 

However, on the algorithmic size, working with instances of large sizes (think of e.g. $N>10^4$), even when we are able to find satisfiable assignments in the clustered region, the BP algorithm initialized in that assignment either converges back to the trivial fixed point, or (in 3-SAT) does not converge at all, in apparent contradiction with the basic assumption. This long-lasting open question had a correspondence in the coloring of random graphs. In the case of coloring it was resolved in \cite{KrzakalaZdeborova10,ZdeborovaKrzakala10} where it was explained using the cavity method that the solutions found by
slow simulated annealing indeed do not have a corresponding BP fixed point, whereas 
% available solving algorithms are not equilibrium ones, and that 
BP initialized in a random solution (obtained via quiet planting) leads indeed to a non-trivial BP fixed point. Whereas one could conjecture that the same reason probably applies to $K$-SAT there was no tool to check this explicitly. The quiet planting that we study in this paper permits such an explicit check for a first time. 

To construct a large random $K$-SAT formula and a non-trivial BP fixed point we proceed as follows: A planted graph at $\lambda^*$ is equivalent to non-planted random graph as long as the constraint density is $\alpha<\alpha_c(\lambda^*)$. This gives us a tool to generate both a typical instance of $K$-SAT and one of its solutions. We then initialize BP from the planted solution, and then iterate the canonical BP equations (non-reweighted, i.e. BP at $\hat\lambda$), we observe that this canonical BP converges to a non-trivial fixed point whenever $\alpha>\alpha_{BP(\hat\lambda)}(\gamma^*)$. From Table \ref{table1} we see that $\alpha_{BP(\hat\lambda)}(\gamma^*)<\alpha_c(\gamma^*)$ for $K\ge 5$. This shows that for $K\ge 5$ we have a non-empty interval of constraint densities where a non-trivial BP fixed point exist on a random formula. Note, however, that in this case (unlike for the coloring) the planted configuration is not a typical configuration with respect to the uniform measure over all solutions $\mu(\hat\lambda)$, therefore the corresponding non-trivial BP fixed point does not describe a random cluster relative to the original, flat, measure. Note also that the fact that for $K>3$ one has $\alpha_{BP(\hat\lambda)}(\gamma^*)< \alpha_d(\hat \lambda)$ is a direct evidence for the presence of small subdominant clusters in the paramagnetic region where a single large cluster still dominates the flat $\mu(\hat\lambda)$ measure.

Freezing of variables was argued to be an important ingredient in understanding the relation between the structure of the space of solutions and the algorithmic hardness of random constraint satisfaction problems \cite{Zdeborova09}. A variable is frozen in a cluster if it takes the same value in all solutions belonging to that cluster. Within the assumptions of the cavity method one can investigate if a solution belongs to a frozen cluster or not by iterating the BP equations initialized in the solution and monitoring whether for some nodes $i$ the BP messages stay completely polarized (i.e. $\chi^i_1=1$ or $\chi^i_0=1$). In fact a simpler version of BP can be written for such monitoring, it is called ``warning propagation'' or ``whitening'' in the literature \cite{Parisi02c,BraunsteinMezard02,BraunsteinZecchina04,ManevaMossel05}. There is a long-standing open question in the field which concerns this whitening \cite{Zdeborova09}. Whereas, for instance, the survey-propagation algorithm is based on the existence of solutions with a non-trivial whitening result, when one tries to run the whitening algorithm on solutions found by heuristic algorithms on large (think e.g. of $N>10^4$) $K$-SAT instances, the whitening result was always observed to be trivial.  
In \cite{ZdeborovaKrzakala10} it was shown for the coloring problem that random solutions do indeed have a non-trivial whitening,  as expected from the theoretical calculations. This same work has also given theoretical reasons expalining why the solutions found by local algorithms do not have a corresponding non-trivial whitening. 
Now we show that random $K$-SAT formulas do have solutions with a non-trivial whitenings. 

Importantly, warning propagation has no dependence on the reweighting $\lambda$, hence if a cluster is frozen at one value of $\lambda$ it must be frozen for all $\lambda$. Table \ref{table1} gives the values of constraint density $\alpha$ beyond which the planted solution lies in a frozen cluster. We use the quiet planting procedure, which generates a typical instance in which the planted solution is also typical, as long as $\alpha <\alpha_c(\gamma^*)$.  We see that, for $K\ge 6$, the planted solution lies in a frozen cluster on typical random instances of $K$-SAT (in the range $\alpha_f(\gamma^*)<\alpha<\alpha_c(\gamma^*)$). 

\begin{figure}[!ht] \centering \includegraphics[scale=0.6]{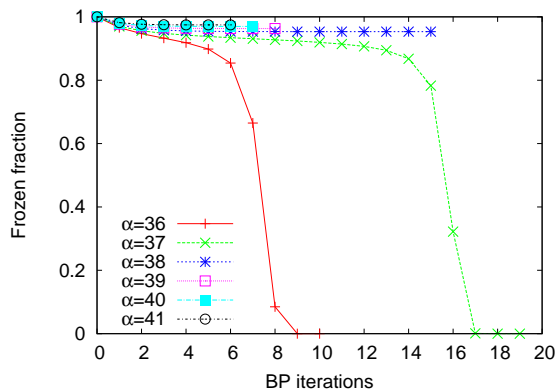} \caption{Non-trivial whitening in the random 6-SAT problem: here $N=10^5$, the instance was created  by planting, but for constraint density where planting provides instances equivalent to random ones. When $\alpha\ge 38$ a non trivial whitening core is obtained for the planted solution. No solution to such a large instance of random $K$-SAT problem with a non-trivial whitening core had ever been found so far.} \label{fig2} \end{figure}

\subsection{Direct evidence for purely entropic barriers}
%Flo - reread this section
%M Je crois en effet qu'il faudrait reprendre un peu cette subsection. Peut-être devrions nous être plus prudents sur le "strong evidence", et utiliser empiricl evidence. Il est vrai que l'objection du referee sur le fait qu'on n'a pas d'anlyse de N fini permettant de montrer que la mesure avec gamma* est bien piégée pendant un temps exponentiel est une objection un peu délicate. On n'a guère envie de refaire des manips, mais si on ne le fait pas il faut trouver une formulation beaucoup plus "qualitative" de nos résultats que ce que nous avions écrit jusqu'à présent.

Often researchers define clusters not via BP fixed points, but as connected components in an auxiliary  graph where vertices are satisfying configurations and edges are drawn between vertices that differ in the value of only one (or some other sublinear in $N$ number of) variable. It has been pointed out on many occasions that there might be important differences between these two definitions. The first definition is more meaningful from the physics point of view, because the BP fixed points are assumed to correspond to basins of attraction of local Monte Carlo Markov chains with detailed balance condition (such MCMC are used to model the behavior of true physical dynamics in materials). A basin of attraction is defined as a part of the configurational space in which the MCMC will stay blocked for a time that grows with the size of the system
%M
 faster than a power law.
Actually, it is often argued that the escape time should grow exponentially with system size $N$, based on a computation of a large deviation function (called the free energy): this function has a minimum corresponding to the cluster (the basin of attraction) in question and all the barriers around this minimum scale linearly with $N$. Since the MCMC acts as a kind of random walk in the landscape given by this free energy function, it takes with high probability an exponential time to find such a path (in the same way it would take exponential time to obtain a configuration on binary variables with more than $60\%$ of ones if one started from a random string of $N$ elements and flipped variables randomly, one by one). 

The barrier in this free energy can have only two possible origins (in general it is a combination of the two)
\begin{itemize}
   \item Energetic barrier: Consider all possible paths from one solution to another (one step being a flip of one variable), if each such path violates at least $B$ clauses at some step, then we say there is an ``energetic barrier" of size $B$ between the two solutions. 
  \item Entropic barrier: Consider an exponentially large subset of all solutions (covering an exponentially small fraction of all solutions), even if the energetic barrier between this subset and most of the other solutions is zero it can still be that a random walk will spend exponential time before escaping from the subset in question. Such a situation is called an entropic barrier.
\end{itemize}

The existence of clusters in the ``energetic" sense, i.e. with the energetic barriers between cluster growing linearly with $N$, was proved for $K$-SAT with large $K$ in \cite{MezardMora05,AchlioptasRicci06}. It is perfectly plausible that clusters with purely entropic barriers exist. Such a situation was described in simple spin glass models (see for instance \cite{BarratMezard95,Ritort95,MoraZdeborova07}), but so far it was not found in a random constraint satisfaction problem. 

Using quiet planting in $K$-SAT we show that in some region of parameters there are purely entropic clusters in random $K$-SAT. 
Consider a planted instance in the regime where the following three conditions are met: there is a planted cluster,  the planting is quiet $\alpha_c(\gamma^*)>\alpha>\alpha_d(\gamma^*)$ and simultaneously the canonical BP converges to a trivial fixed point $\alpha<\alpha_{BP(\hat \lambda)}(\gamma^*)$ (for values see Table~\ref{table1}). Recall that the set of satisfying configurations (``solutions'') is independent of the reweighting $\lambda$. If we initialize the BP reweighted by $\gamma^*$ in the planted configuration the iterations will converge to a non-trivial fixed point, meaning that there is a planted cluster in the measure reweighted $\mu(\gamma^*)$. If we run canonical non-reweighted BP the iterations will converge to the factorized fixed points (from both initializations - in the planted configurations and in the fixed point of the reweighted BP), meaning that there is no planted cluster in the flat measure $\mu(\hat\lambda)$. Hence the dynamics of non-reweighted BP was able to escape from the region that formed a cluster in the measure $\mu(\gamma^*)$. 

Relying on the conjectured relation between BP fixed points and basins of attraction of Monte Carlo Markov chains the above result translates into the following numerical experiment with a MCMC random walk in the space of solutions. 

\begin{figure}[!ht] \centering \includegraphics[scale=0.6]{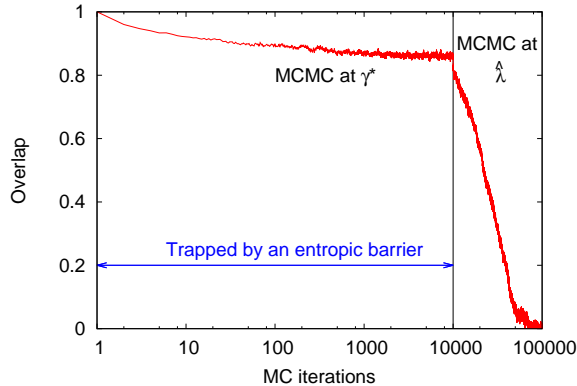} \caption{Illustration of the existence of purely entropic barriers: a solution is planted using $\gamma^*$, $\alpha=8.2$, $N=10^5$, $K=4$. Then a Monte-Carlo simulation satisfying detailed balance for the measure $ \mu(\lambda^*)$ is performed starting from this planted configuration: the system is trapped in a cluster and does not exit it for a very long time. Switching then to a new Monte Carlo, which satisfied detailed balance for the measure $\mu(\hat\lambda)$, i.e. the flat measure over all solutions, makes it very easy for the MCMC to exit this cluster and decorrelate from the initial solution. This gives a strong evidence for the presence of entropic barriers and the fact that the dynamics can be trapped even in absence of frozen variables or energetic barriers.
% and b) the presence of smaller Gibbs states in the paramagnetic region where a large single Gibbs state still dominates the $\mu(\hat\lambda)$ measure.
}
\label{fig4}  
\end{figure}

First, initialize in the planted configuration, consider a MCMC random walk that satisfies the detailed balance condition with respect to the measure $\mu(\lambda^*)$. This random walk will be restrained in proximity of the planted configuration and stay there for a very long time. We conjecture that this time is exponential in the size of the system based on the possible calculation of the barrier in the free energy and the relation between clusters and fixed points of BP equations. We do not attempt to have a full numerical confirmation of such exponential scaling: in the numerical experiment of Fig.~\ref{fig4} we content ourselves with a time long enough to see a saturation of the overlap with the initial configuration in the logarithmic time-scale.  

In a second part of this numerical experiment, consider a Monte Carlo that satisfies the detailed balance  with respect to the $\mu(\hat\lambda)$ measure. This MCMC will be able to walk far away from the planted configuration in relatively short time. This means that it  manages to find some paths that were very rare in the measure $\mu(\lambda^*)$ but are now easy to find  in $\mu(\hat\lambda)$. Using such paths, the second procedure escapes to a large distance from the planted configuration. These experiments are illustrated in Fig.~\ref{fig4}. This provides a strong empirical evidence for the existence of purely entropic barriers.

\subsection{How hard is $\lambda^*$-planted $K$-SAT}
\label{sec:hard}
Hardness of the $\gamma^*$-planted (we remind that $\gamma^*$ is only a special point in the space of $\lambda^*$) random $3$-SAT instances was investigated in  \cite{MooreJia05}  for DPLL, WalkSAT and Survey Propagation algorithm. The authors found empirically a region where all the three algorithms fail or scale exponentially with the size of the system. Our results actually show that for simulated annealing or for belief propagation decimation random $\gamma^*$-planted $3$-SAT is easy. Indeed, for $\alpha>\alpha_l(\gamma^*)$ the BP and MCMC dynamics is attracted close to the planted configuration and hence finding a solution nearby is easy. For $\alpha<\alpha_l(\gamma^*)$ the planted formula is equivalent to a random formula, and it is so sparse that MCMC will equilibrate in linear time and hence simulated annealing will find a solution in linear time, in this region also BP decimation works \cite{MontanariRicci07,RicciSemerjian09}. 
On the other hand for $K\ge 4$ the $\gamma^*$-planted $K$-SAT has a hard region for values of $\alpha_s < \alpha < \alpha_l(\gamma^*)$. Indeed for  $\alpha < \alpha_l(\gamma^*)$ the planted cluster is hidden to the MCMC and BP dynamics and for $\alpha>\alpha_s$ there are no solutions other than those belonging to the planted cluster. At this point we should mention that whereas we do not make any claim about hardness for an arbitrary algorithm, the class of algorithms for which the planted cluster is hard to find for $\alpha>\alpha_s$ seems to be quite large, it includes the local search algorithms, as well as message-passing ones. But for instance it would not include algorithms such as Gaussian elimination for constraint satisfaction problems that are linear over some field, such as the $K$-XOR-SAT problem.

The algorithmic hardness of $\lambda^*$-planted 3-SAT formulas was studied in \cite{BarthelHartmann02} using a walk-SAT algorithm. We revisit the corresponding phase diagram in view of our results. We show that the conclusions of \cite{BarthelHartmann02} were qualitatively correct, but the correct boundary of the ``hard" region is different from what was estimated in \cite{BarthelHartmann02} (as the statistical physics calculations in that work were only approximate). Most importantly we give further theoretical justification for why the $\lambda^*$-planted random 3-SAT formulas are (together with benchmarks from \cite{HaanpaaJarvisalo06}) the hardest satisfiable formulas on the market of hard satisfiable benchmarks. 

We remind that the space of solution in the $\lambda^*$-planted formulas can be split in two parts. The first part includes all the solutions (satisfying assignments) that would exist anyway in the non-planted random 3-SAT.  The second part includes solutions correlated to the planted configuration. Algorithmic search for a solution belonging to the first part is just as hard (or as easy) as it is in the canonical random $K$-SAT. Notably for $\alpha>\alpha_s$ this space is empty (canonical random SAT is unsatisfiable). It follows from our results above that for $\alpha>\alpha_l$ it is easy to find a solution correlated with the planted configuration, to do so one can use $\lambda^*$-reweighted BP or $\lambda^*$-reweighted MCMC. We argue that, for $\alpha<\alpha_l$, finding solutions correlated to the planted one is hard. Finding a configuration (not necessarily a solution) correlated to the $\lambda^*$-planted configuration may be seen as an inference problem and running $\lambda^*$-reweighted MCMC is then a Bayes-optimal algorithm for this inference problem. The cavity method predicts that a barrier of size linear in the system size would have to be overcome in order to find a configuration correlated to the planted one. Hence the running time of the MCMC would become exponentially large below the transition $\alpha_l$ (and above $\alpha_d$). Concerning algorithms such as BP, the metastable state in which MCMC is blocked corresponds to a stable BP fixed point and one would have to initialize the BP messages extremely close to the planted configuration to find another fixed point. With random initializations this event would be exponentially rare. Clearly if it is exponentially hard to find any {\it configuration} correlated to the planted configuration, it will be even harder to find a {\it solution} correlated to the planted configuration. To conclude, the boundary of the hard region is given by $\alpha<\alpha_l$ in order not to find the planted configuration, and at the same time $\alpha$ sufficiently large for the canonical random $K$-SAT to be hard or better not to have any solutions at all for $\alpha>\alpha_s$. This region is depicted by shading in Fig.~\ref{fig0} for 3-SAT.   

Let us also describe explicitly the relation between the benchmarks described here and those of \cite{HaanpaaJarvisalo06}. Instances generated by the $\lambda^*$-planting with $\lambda_2^*$ close to zero can be seen as a planted XOR-SAT formulas with ``a bit of nonlinearity''. If moreover we restrict ourselves to the regular formulas where every variable appears in the same number of clauses then we are exactly in the setting of  \cite{HaanpaaJarvisalo06} who introduced planted regular XOR-SAT formulas with ``a bit of nonlinearity'' as the hardest known satisfiable benchmarks. Moreover the kind of nonlinearity that quiet planting adds by taking a nonzero value of $\lambda_2^*$ is not easy to discover, even if the exact protocol of how the instances were created was known.  Note that the regular instances are taken in order to lower the total number of variables (e.g. the leaves of random instances do not contribute to the overall hardness and can hence be omitted). The analysis presented in this paper can be used for regular instances straightforwardly.

\section{Conclusions}
In this paper, we have studied the reweighted measure over solutions of the random $K$-satisfiability problem, where each solution is reweighted according to the number of variables that satisfy every clause. This problem has been addressed both analytically, using the cavity method, and numerically, using belief propagation and Monte-Carlo Markov chains. 

The main results of this study are the following: (i) The reweighting allows to introduce a planted ensemble that generates instances that are equivalent to random instances in some region of the phase diagram. In this case, we are thus able to generate simultaneously a typical SAT instance and one of its solutions. (ii) We have used this property to give a strong evidence for the existence of purely entropic (rather than energetic) barriers between clusters in some region of the phase diagram. This is a fundamental point demonstrating that the physical definition of clusters as basins of attraction (or belief propagation fixed points) is {\it not} equivalent to the geometric definition of clusters as  disconnected components in the solution space. Such equivalence was often wrongly assumed in the literature and leads to confusions.
(iii) We have used the quiet planting property to display explicitly, for the first time, solutions of large random K-SAT problems leading to a non-trivial whitening core: while such solutions where known to exist, they were so far never observed on very large instances. (iv) We discuss the algorithmic hardness of these planted instances and determine a region of parameter in which quiet planting leads to hard satisfiable benchmarks.

\bibliographystyle{plainbv}
\bibliography{myentries}

\end{document}